# The force as a function: Towards analytical graphic statics for spatial structures

Tamás BARANYAI*

*Budapest University of Technology and Economics
Hungary, 1111 Budapest, Műegyetem rakpart 1-3. K.II.61.
baranyai.tamas@epk.bme.hu

## Abstract

One of the most influential early works in graphic statics is one of Maxwell, where he introduced the idea of a discontinuous stress function and the use of a 3D projective polarity for a planar problem. A recent work gave an analytical description with the goal to provide explaining power, treating planar forces as linear functionals (moment functionals) forming a 3D vector space, so force and form diagrams of planar problems can be interpreted in a 3-dimensional way. The linear combination of these moment functions can naturally be used as a discontinuous stress function since the Airy stress function is known to correspond to moments of planar forces. The main contribution of this work is to present a dimension independent way of treating forces as functions, that returns the known stress-functions of planar and spatial graphic statics. This is done by relying on the multi-linear function definition of Grassmann-algebra.

## 1. Introduction

In an often-cited paper (Maxwell [8]) Maxwell showed how given two 3 dimensional polyhedra dual to each other with respect to a paraboloid of revolution, the projection of the edges of the polyhedra parallel to the axis of revolution to an orthogonal (to the axis) image plane will give two diagrams reciprocal to each other. This reciprocity is the property that if one diagram is considered as a geometry of a (statically indeterminate) truss, the other diagram will be its force-plan, representing the magnitude of each force with a line segment orthogonal to said force, with length proportional to the magnitude of the force.

Maxwell also noted, how the planes constituting the polyhedron over the form diagram can be interpreted as a discrete analogue of what is now known as Airy-stress function. This aspect was heavily focused on in recent works (Mitchel et. al. [12], McRobie at. al. [10]) on graphic statics noting how for beams the moment diagram of the beam also appears under these surfaces (Hegedűs [5], Williams and McRobie [18]). The possibility to use the moment function as a stress function is an older result from the 1930's, (Phillips [13]), which was rediscovered by the author when providing an analytical description of the projective duality involved (Baranyai [1]). This work serves as a continuation of the previous in aiming for an analytical description spatial graphic statics. As past works treated planar and spatial graphic statics somewhat differently, a dimension independent description is aimed at. We will take a brief look at said difference below, to explain this.

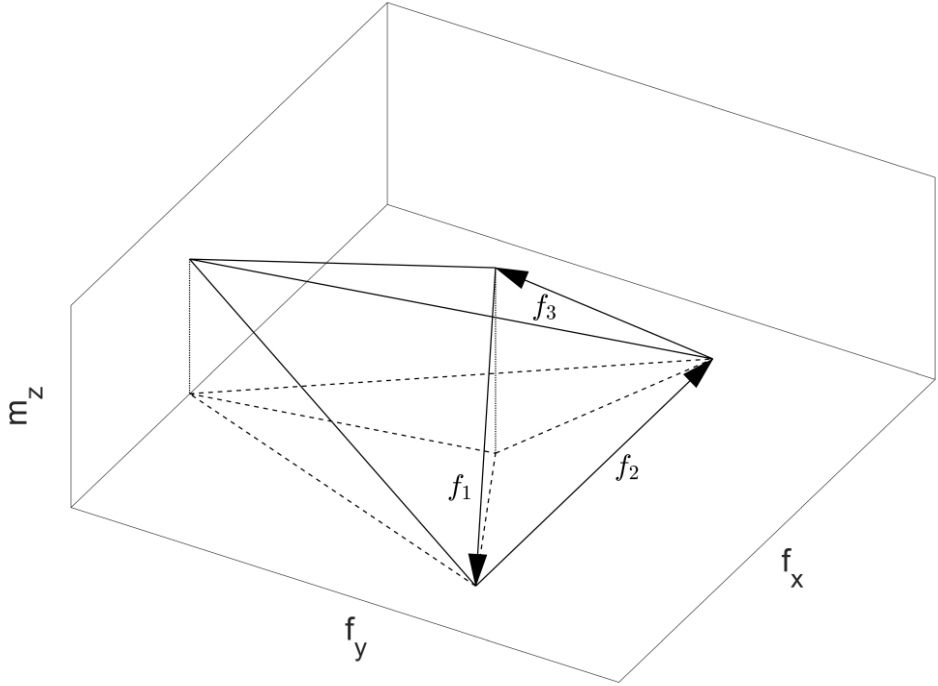

1. Figure Three-dimensional force diagram for planar graphic statics. The two-dimensional old diagram is present in the projection. The geometry of the corresponding truss can be seen in Figure 2.

We will use the convention that whenever there is an engineering problem involved, we will use an orthogonal coordinate system and label its directions with $x,y$ and $z$. When using projective homogeneous coordinates, we will number the axes for example $x_0 \ldots x_{n+1}$.

The gist of planar graphic statics can be seen in the ability to represent a force $s_i$ both with a vector and a function. The vector can be given as $\boldsymbol{f_i} = (f_x, f_y, m_z)$ where $f_x$ and $f_y$ are the components of the force and $m_z$ is the moment of the force with respect to the origin of the coordinate system chosen. Each equilibrium equation can be considered as an arrow continuous loop of vectorial addition in $\mathbb{R}^3$. An example to this three-dimensional force diagram is presented in Figure (1). The function can be given as

$$\varphi(x, y) = -f_y x + f_x y + m_0 \qquad (1)$$

which is the moment of the force with respect to a point in the plane. The equilibrium in this case can be seen by the resultant being the constant 0 function. This is illustrated in Figure 2.

When embedding the problem into the projective plane equation (1) can be thought of as the evaluation of a linear functional $\boldsymbol{\varphi}$ over the subspace $(x, y, 1)$, which is a set of appropriately chosen homogenous point-coordinates. The duality of vectors and functionals is tied to the projective duality concept and Maxwell's three-dimensional duality nicely corresponds to the algebraic three-dimensionality of the problem at hand. On the other hand, for spatial graphic statics there is a bivector based stress function (McRobie and Williams [11]), built up from vectors thought of as plane representants. The duality used for these graphic constructions is 4 dimensional (McRobie [9]), instead of the six-dimensional space of forces and moments.

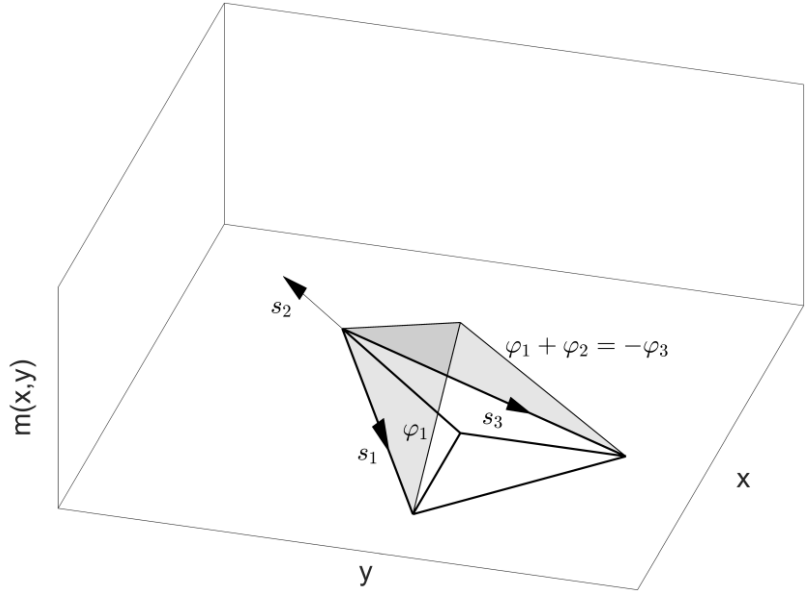

2. Figure Illustration of the function interpretation of Maxwells construction. The three-dimensional force diagram corresponding to this truss can be seen in Figure 1.

## 2. Preliminaries

In screw theory dealing with both forces and velocities of bodies, forces (static dynames) are often called co-screws (Selig [15]): linear functionals over the velocities (screws). In statics we have no velocities, but we still like to think about forces as functions over geometric objects, that would correspond to some motion the reaction forces do not allow. In $\mathcal{P}^d$ a rigid body motion corresponding to a "simple" screw has a pointwise fixed projective subspace of dimension $d-2$, meaning a static dyname can be thought of as a function over $d-2$ dimensional projective subspaces. As such subspaces can be represented both as an intersection of two $d-1$ dimensional subspaces and the span of $d-1$ points, the corresponding description can be either a bilinear function or a $(d-1)$-linear function. A high-level view of the duality transformations in graphic statics is the interchange of these two representations. Since both screws and co-screws are elements of finite dimensional real vector-spaces that are isomorphic to each other, the choice which is the vector and which is the function is the choice of the user and is application dependant. As we will go into the details below, we will adopt a geometry motivated vector-function choice and we will "build" our forces from points (vectors) or hyperplanes (functions). For algebraic purposes this distinction is not strictly necessary, it is made to reflect on the duality based graphic methods of graphic statics.

### 2.1. Notation

Here the used notation and concepts are briefly defined, partially due to the fact, that different authors use concepts of Grassman-algebra slightly differently. The main sources for this part were (Szőkefalvi *et al.* [16]) and (Marsh [7]).

*2.1.1 Parts of Grassmann algebra*

Vectors will be denoted with bold while scalars with italics. We will differentiate with Latin and Greek letters between primal and dual vector-spaces. Let us denote a base in $\mathbb{R}^n$ with $\{\boldsymbol{e}_1, \boldsymbol{e}_2, \dots \boldsymbol{e}_n\}$, while a base in $(\mathbb{R}^n)^*$ (the space of linear functionals) with $\{\boldsymbol{\varepsilon}_1, \boldsymbol{\varepsilon}_2, \dots \boldsymbol{\varepsilon}_n\}$.

A k-form over $\mathbb{R}^n$ is a map $\boldsymbol{\Omega}: \mathbb{R}^{nk} \to \mathbb{R}$ satisfying:

1. Linearity in all $k$ vector-variables
2. If two of the vectors $\boldsymbol{v}_1, \dots \boldsymbol{v}_k$ are the same, then $\boldsymbol{\Omega}(\boldsymbol{v}_1, \dots \boldsymbol{v}_k) = 0$.

The second property is equivalent with the map being alternating, that is transposing two vectors in the input negates the sign of the result. The vector-space of these k-forms is known to be $\binom{n}{k}$ dimensional, and one can construct such maps as an exterior product of linear functions $\boldsymbol{\Omega} = \boldsymbol{\omega}_1 \wedge \boldsymbol{\omega}_2 \wedge \dots \wedge \boldsymbol{\omega}_k$ using the evaluation-rule:

$$(\boldsymbol{\omega}_1 \wedge \boldsymbol{\omega}_2 \wedge \dots \wedge \boldsymbol{\omega}_k)(\boldsymbol{v}_1, \boldsymbol{v}_2, \dots, \boldsymbol{v}_k) = \det|\boldsymbol{\omega}_i(\boldsymbol{v}_j)|. \tag{2}$$

In some books (Szőkefalvi *et al*. [16]) this is multiplied with $\frac{1}{k!}$, which we dropped due to the mechanical interpretation presented in equation (5).

Given a base in $(\mathbb{R}^n)^*$ one can have the Hodge-dual of k-forms, which we will give on the base. Let $\Pi$ be an ordered index set from indices $\{1,2,\dots n\}$ and $(\boldsymbol{\varepsilon}_{\Pi(1)} \wedge \dots \wedge \boldsymbol{\varepsilon}_{\Pi(k)})$ a k-form! The Hodge-dual (denoted by $*(\dots)$) of this form is the $n-k$ form

$$* (\boldsymbol{\varepsilon}_{\Pi(1)} \wedge \dots \wedge \boldsymbol{\varepsilon}_{\Pi(k)}) = (-1)^t (\boldsymbol{\varepsilon}_{\Pi(k+1)} \wedge \dots \wedge \boldsymbol{\varepsilon}_{\Pi(n)}) \tag{3}$$

where $t$ is the number of transpositions required to bring $\Pi$ to $(1,2,\dots n)$. We can note this is an isomorphism between $\wedge^k (\mathbb{R}^n)^*$ and $\wedge^{n-k} (\mathbb{R}^n)^*$, possible due to $\binom{n}{k} = \binom{n}{n-k}$. As $\mathbb{R}^n$ and $(\mathbb{R}^n)^*$ are also isomorphic we can also have exterior products of vectors defined in a similar way. Furthermore, we can combine the natural isomorphism between $\mathbb{R}^n$ and $(\mathbb{R}^n)^*$ with the Hodge-duality to get a map between $k$-forms and $(n-k)$-vectors. It will be denoted with $\dagger(\dots)$ and is given as

$$\dagger (\boldsymbol{\varepsilon}_{\Pi(1)} \wedge \dots \wedge \boldsymbol{\varepsilon}_{\Pi(k)}) = (-1)^t (\boldsymbol{e}_{\Pi(k+1)} \wedge \dots \wedge \boldsymbol{e}_{\Pi(n)}). \tag{4}$$

*2.1.2 Projective homogeneous coordinates*

Consider $\mathbb{R}^{d+1}$ with the equivalence relation $\boldsymbol{v} \sim \alpha \boldsymbol{v} \mid \alpha \neq 0$ and $\boldsymbol{0} \neq \boldsymbol{v} \in \mathbb{R}^{d+1}$! Factorizing $\mathbb{R}^{d+1} / \{0\}$ with this relation gives a coordinate description of $d$ dimensional projective space $\mathcal{P}^d$. In essence a $k$ dimensional projective subspace is identified with a $k+1$ dimensional linear subspace. We will (unless otherwise indicated) follow the convention that points in the Euclidean part of $\mathcal{P}^d$ have a representant of form $(\underline{\boldsymbol{v}}, 1)$ with $\underline{\boldsymbol{v}} \in \mathbb{R}^d$ while ideal points (points at infinity) have representants of form $(\underline{\boldsymbol{v}}, 0)$. Projective subspaces of dimension $d-1$ can be represented with equivalence classes of linear functions that vanish over their point representants: $\boldsymbol{v}$ represents a point in subspace $\boldsymbol{\sigma}$ exactly if $\boldsymbol{\sigma}(\boldsymbol{v}) = 0$.

*2.1.3 The Minkowski problem*

We will also make use of the solution to the Minkowski problem (Klain [6]), also known as Minkowski's theorem:

Theorem 1: Suppose $\boldsymbol{u}_1, \boldsymbol{u}_2, \dots, \boldsymbol{u}_k \in \mathbb{R}^n$ are unit vectors that span $\mathbb{R}^n$; and suppose that $\alpha_1, \alpha_2, \dots, \alpha_k > 0$. Then there exists a convex polytope in $\mathbb{R}^n$; having facet unit normals $\boldsymbol{u}_1, \boldsymbol{u}_2, \dots, \boldsymbol{u}_k$ and corresponding facet areas $\alpha_1, \alpha_2, \dots, \alpha_k$ if and only if: $\alpha_1 \boldsymbol{u}_1 + \alpha_2 \boldsymbol{u}_2 + \dots + \alpha_k \boldsymbol{u}_k = 0$. Moreover, this polytope is unique up to translation.

## 3. Forces as bivectors

Consider force $\underline{f} \in \mathbb{R}^d$ acting at point $\underline{p} \in \mathbb{R}^d$, with respective homogenous coordinate vectors $\boldsymbol{f} = \left(\underline{f}, 0\right)$ and $\boldsymbol{p} = \left(\underline{p}, 1\right)$. The bivector representing it can be given as $\boldsymbol{p} \wedge \boldsymbol{f}$ (see for instance (Whiteley [17])). In case of a pure moment one can choose a point at infinity as $\boldsymbol{p} = \left(\underline{p}, 0\right)$ with $\|\underline{p}\| = 1$.

Consider two distinct $d-1$ dimensional projective hyperplanes represented by homogeneous coordinate vectors $\boldsymbol{\xi_1} = \left(\underline{x_1}, -d_1\right)$ and $\boldsymbol{\xi_2} = \left(\underline{x_2}, -d_2\right)$. Depending on whether the intersection $(\boldsymbol{\xi_1} \wedge \boldsymbol{\xi_2})$ represents a subspace in the ideal of affine part of $\mathcal{P}^d$ evaluating this bilinear form over the vector pair $(\boldsymbol{p}, \boldsymbol{f})$ can be considered the scalar multiple of the projection of the force, or the moment of the force with respect to the subspace. To have exactly the right value one has to appropriately pick the representants of $\boldsymbol{\xi_1}$ and $\boldsymbol{\xi_2}$. One way of doing it is as follows: For the moment equation prescribe $\underline{x_1} \perp \underline{x_2}$ and $\|\underline{x_1}\| = 1 = \|\underline{x_2}\|$. This way the evaluation is

$$(\boldsymbol{\xi_1} \wedge \boldsymbol{\xi_2})(\boldsymbol{p}, \boldsymbol{f}) = \begin{vmatrix} \langle \underline{x_1}, \underline{p}\rangle - d_1 & \langle \underline{x_2}, \underline{p}\rangle - d_2 \\ \langle \underline{x_1}, \underline{f}\rangle & \langle \underline{x_2}, \underline{f}\rangle \end{vmatrix} = \left(\langle \underline{x_1}, \underline{p}\rangle - d_1\right)\langle \underline{x_2}, \underline{f}\rangle - \left(\langle \underline{x_2}, \underline{p}\rangle - d_2\right)\langle \underline{x_1}, \underline{f}\rangle \quad (5)$$

where $\langle , \rangle$ represents the usual scalar product. After remembering that $d_i$ are the respective distances of the hyperplanes from the origin and vectors $\underline{x_1}$ and $\underline{x_2}$ serve as two successive elements of an orthonormal base it can be seen to be the required moment value (see Figure (3)).

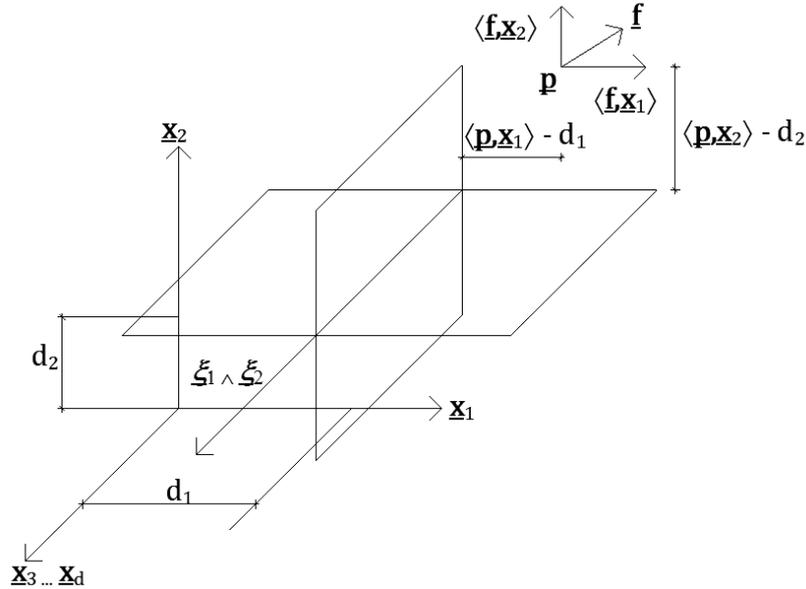

3. Figure: Illustration of the moment equation. Note how this is a d dimensional figure with axes $x_3 \ldots x_d$ projected into one.

For the projection of the force prescribe $\|\underline{x_1}\| = 1$, $\|\underline{x_2}\| = 0$ and $d_2 = 1$. This will give the projection of the force in the $\underline{x_1}$ direction. Note, how in this case $\boldsymbol{x_2}$ represents the hyperplane at infinity.

It might be beneficial to compare this with kinematics and recall that $(\boldsymbol{\xi_1} \wedge \boldsymbol{\xi_2})$ can be thought of as an instantaneous rotation (Gunn [4]) around the pointwise fix projective subspace $(\boldsymbol{\xi_1} \wedge \boldsymbol{\xi_2})$ represents.

## 4. Forces as multilinear functions

As the duality introduced in equation (4) suggests, one can take the dual of the bivector formulation of section 2.2 and get static dynames "built" from projective hyperplane coordinates functioning as alternating multilinear functions over $d-1$ vectors representing multiple points of $\mathcal{P}^d$. It will be shown below how this will give the known stress functions of planar and spatial graphic statics.

### 4.1. $\mathcal{P}^2$

Let us pick an orthonormal coordinate system in the plane and label its axes $x$ and $y$! Let us embed the Euclidean space into $\mathcal{P}^2$ such that Euclidean points have a representant of form $\boldsymbol{p} = (p_x, p_y, 1)$ while the force will "correspond" to vector $\boldsymbol{f} = (f_x, f_y, 0)$! With this the effect of duality equation (4) can be coordinate-wise described as

$$-f_y(\boldsymbol{e}_2 \wedge \boldsymbol{e}_3) \mapsto -f_y \boldsymbol{\varepsilon}_1 \tag{6}$$

$$-f_x(\boldsymbol{e}_1 \wedge \boldsymbol{e}_3) \mapsto f_x \boldsymbol{\varepsilon}_2 \tag{7}$$

$$(p_x f_y - p_y f_x)(\boldsymbol{e}_1 \wedge \boldsymbol{e}_2) \mapsto (p_x f_y - p_y f_x)\boldsymbol{\varepsilon}_3. \tag{8}$$

The right sides constitute a component-wise description of a 1-form, the moment functional given in (Baranyai [1]).

### 4.2. $\mathcal{P}^3$

To be consistent with previous works on spatial graphic statics, in the 3-dimensional case we will label the base vectors of $\mathbb{R}^4$ as $\boldsymbol{e}_0, \cdots, \boldsymbol{e}_3$, while base vectors of $(\mathbb{R}^4)^*$ as $\boldsymbol{\varepsilon}_0, \cdots, \boldsymbol{\varepsilon}_3$. Let us embed our Euclidean points as $\boldsymbol{p} = (1, p_x, p_y, p_z)$ while the force will take the form of $\boldsymbol{f} = (0, f_x, f_y, f_z)$. As a notational shorthand let us introduce the moment vector $(m_x, m_y, m_z) = (p_x, p_y, p_z) \times (f_x, f_y, f_z)$. With this the component-wise description of the duality given in equation (4) is:

$$f_x(\boldsymbol{e}_0 \wedge \boldsymbol{e}_1) \mapsto f_x(\boldsymbol{\varepsilon}_2 \wedge \boldsymbol{\varepsilon}_3) \tag{9}$$

$$f_y(\boldsymbol{e}_0 \wedge \boldsymbol{e}_2) \mapsto -f_y(\boldsymbol{\varepsilon}_1 \wedge \boldsymbol{\varepsilon}_3) \tag{10}$$

$$f_z(\boldsymbol{e}_0 \wedge \boldsymbol{e}_3) \mapsto f_z(\boldsymbol{\varepsilon}_1 \wedge \boldsymbol{\varepsilon}_2) \tag{11}$$

$$m_z(\boldsymbol{e}_1 \wedge \boldsymbol{e}_2) \mapsto m_z(\boldsymbol{\varepsilon}_0 \wedge \boldsymbol{\varepsilon}_3) \tag{12}$$

$$-m_y(\boldsymbol{e}_1 \wedge \boldsymbol{e}_3) \mapsto m_y(\boldsymbol{\varepsilon}_0 \wedge \boldsymbol{\varepsilon}_2) \tag{13}$$

$$m_x(\boldsymbol{e}_2 \wedge \boldsymbol{e}_3) \mapsto m_x(\boldsymbol{\varepsilon}_0 \wedge \boldsymbol{\varepsilon}_1) \tag{14}$$

We will compare this with a previous work on spatial graphic statics by McRobie and Williams (McRobie and Williams [11]). According to equation (2) in it, forces are calculated from linear functions $\boldsymbol{\gamma}_i = (\gamma_{i,0}, \underline{\boldsymbol{\gamma}}_i)$ as

$$\boldsymbol{F} = \tfrac{1}{2}\sum_i \underline{\boldsymbol{\gamma}}_i \times \underline{\boldsymbol{\gamma}}_{i+1} \tag{15}$$

where $i$ runs on a cyclical index set. (The Greek letter notation has been forced on the equation by the author of this paper.) As the cross product negates the minus sign on the right side of equation (10), it can be seen how equation (15) is a linear combination of the base vectors $(\boldsymbol{\varepsilon}_i \wedge \boldsymbol{\varepsilon}_j)$ from the right sides of equations (9-11). The referenced paper then explains how it uses the other 3 components to represent the moments and the stress function is given as (equation 29, forced into Greek):

$$\boldsymbol{R} = \tfrac{1}{2}\sum_i \boldsymbol{\gamma}_i \wedge \boldsymbol{\gamma}_{i+1}. \tag{16}$$

One could dualize the evaluation rule in equation (5) and evaluate $\boldsymbol{R}$ over a pair of vectors $(\boldsymbol{x}_1, \boldsymbol{x}_2)$ representing a pair of points.

## 5. Forces as oriented *n-1* volumes

There are several representational ideas for different graphic static problems, originating from Maxwell (Maxwell, [8]) and Cremona (Cremona [2]) for planar problems and from Rankine (Rankine [14]) for spatial ones. The latter can be considered a generalization of the planar concept of Maxwell into 3 dimensions, and the method generalizes further, which has been already noted before (Whitely, [17]). A force is represented this way with a $d-1$ dimensional polytope orthogonal to the represented force, with the $d-1$ dimensional volume of the polytope corresponding to the magnitude of the force. This is the use of the Minkowski-theorem and the paper will refer to such constructions as Minkowski reciprocals in the following (even in lower dimensions).

The exterior product of $d-1$ functionals can be readily interpreted as an oriented $d-1$ volume, an interpretation which we borrow from geometric algebra. For $d=3$ this is proposed in (McRobie [9] and McRobie and Williams [18]), the former also explains why the $d-1$ polytope need not contain the origin. At this point let us note, that since forces have been identified with ideal points in section 3, the force components of the static dynames will correspond to bivectors of form $(\boldsymbol{e}_i \wedge \boldsymbol{e}_d)$ (or $(\boldsymbol{e}_0 \wedge \boldsymbol{e}_i)$ with the convention of section 4.2) for some $i$. In $(\mathbb{R}^{d+1})^*$ these components correspond to polytopes not having volumes in the $\boldsymbol{\varepsilon}_d$ (or $\boldsymbol{\varepsilon}_0$) direction. As such, if we project a polytope parallel to this direction in $(\mathbb{R}^{d+1})^*$, the force component is left unchanged. It is this projection that has been used when projecting a (3D or 4D) polytope to get a Minkowski force diagram. Here we gave some detail on how this interpretation generalizes in dimensions.

Although these polytopes inherently exist in $(\mathbb{R}^{d+1})^*$, this space can be identified with the Euclidean part of $\mathcal{P}^{d+1}$ meaning one can arrive to these polytopes (the set of points spanning them) through a projective duality from a set of hyperplanes. The actual duality used may vary, depending on how one defines the planes dual to the points of the aforementioned $d+1$ polytopes. It seems to be natural to similarly identify $\mathbb{R}^{d+1}$ with the Euclidean part of $\mathcal{P}^{d+1}$ and thus draw the representant choice of homogenous point coordinates (this is not the same construction Maxwell used). The process is explained in the next section.

## 6. Minkowski reciprocals of trusses using the presented duality

As not all structures admit this type of force-diagram and we are interested in the projective geometric duality used to get such force-diagram, we assume it exists and start with the higher dimensional polyhedron. Let $\mathcal{C}_g \subset \mathbb{R}^{d+1}$ be a convex, $d+1$ dimensional polyhedron, and the projection of its edges to the $x_{d+1} = 1$ hyperplane the $d$ dimensional truss. The projection is central through the origin. Algebraically speaking this means if a force in the $d$ dimensional problem is given by the bivector $(\underline{\boldsymbol{p}}, 1) \wedge (\underline{\boldsymbol{f}}, 0)$ (where $\underline{\boldsymbol{p}}, \underline{\boldsymbol{f}} \in \mathbb{R}^d$) then there are scalars $\alpha_p, \alpha_f$ such that points of an edge of $\mathcal{C}_g$ are given by either

$$\alpha_p \beta_p (\underline{\boldsymbol{p}}, 1) + \alpha_f \beta_f (\underline{\boldsymbol{f}}, 0) \mid \beta_p + \beta_f = 1 \tag{17}$$

or

$$\alpha_p \beta_p (\underline{\boldsymbol{p}}, 1) + \alpha_f \beta_f (\underline{\boldsymbol{f}}, 0) \mid \beta_p = 1. \tag{18}$$

Equation (17) represents the case where the line is going through $\alpha_p (\underline{\boldsymbol{p}}, 1)$ and $\alpha_f (\underline{\boldsymbol{f}}, 0)$ while equation (18) the case where the edge is parallel with the plane $x_{d+1} = 1$. In both cases the effect of $\alpha_p$ and $\alpha_f$ can be considered as choosing alternative representants of $d$ dimensional homogeneous coordinates.

This construction relies on the projection inherently present in the equivalence classes of homogeneous coordinates, instead of creating another one. It can be seen in Figure (4).

Embedding the problem into $\mathcal{P}^{d+1}$ and taking the canonical dual of the line gives a subspace whose finite points $q_i$ have representants $\left(\underline{q_i}, \delta_i, 1\right)$ with some $\delta_i \in \mathbb{R}$ that satisfy

$$\alpha_p \langle \underline{p}, \underline{q_i} \rangle + \alpha_p \delta_i + 1 = 0 \qquad (19)$$

and

$$\alpha_p \langle \underline{f}, \underline{q_i} \rangle + \alpha_p 0 + 1 = 0. \qquad (20)$$

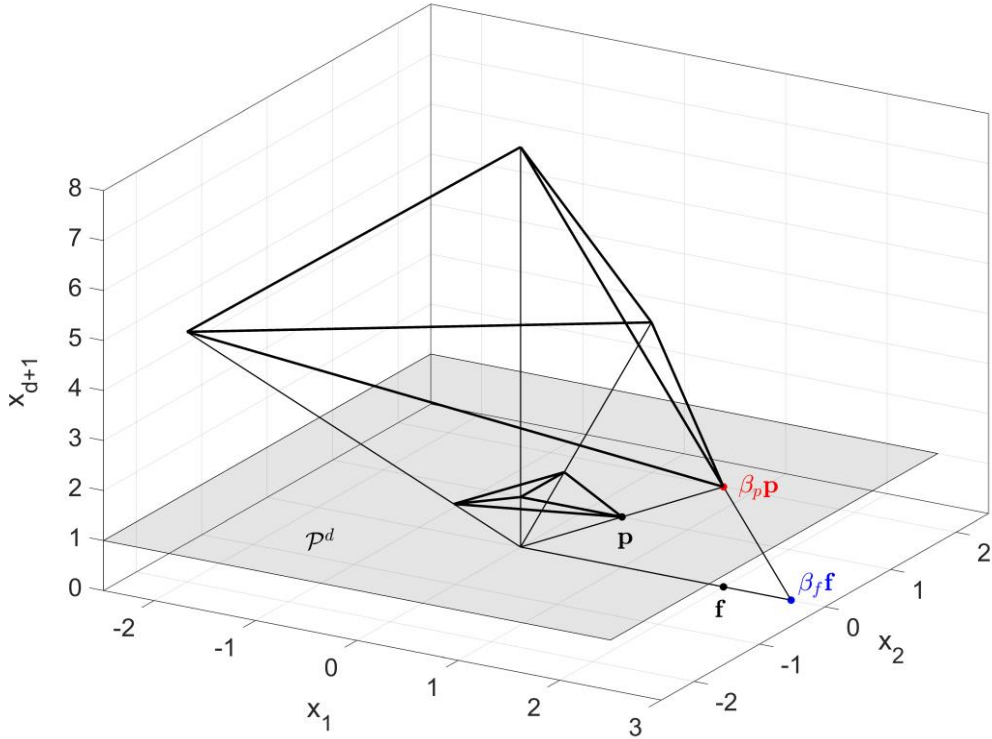

4. Figure: Geometric interpretation of the representant choice in $\mathbb{R}^{d+1}$

As the $\mathcal{C}_g$ was convex its dual polyhedron $\mathcal{C}_f$ will have the properties we usually associate with dual polyhedra (Grünbaum and Shephard [3]) and it will have a convex face (dual to the edge of $\mathcal{C}_g$) lying in the subspace defined by (19) and (20). Projecting down this face parallel with $e_{d+1}$ to the $x_{d+1} = 1$ subspace (which is the Euclidean part of $\mathcal{P}^d$) will give $d$ dimensional projective homogeneous coordinates as $\left(\underline{q_i} 1\right)$. It can be easily checked from equation (20) how for any pair of points $\left(\underline{q_i}, \delta_i, 1\right)$ and $\left(\underline{q_j}, \delta_j, 1\right)$ of this face their projections have representants $\left(\underline{q_i}, 0, 1\right)$ and $\left(\underline{q_j}, 0, 1\right)$ and the difference $\underline{q_i} - \underline{q_j}$ is orthogonal to $\underline{f}$. As this holds for any pair of points the projection of the face is orthogonal to $\underline{f}$. This, together with the topological properties guaranteed by the convexity and the duality means the projection of $\mathcal{C}_f$ has the properties to represent the forces of the truss through Minkowski's theorem.

The full details of this construction are missing, as an interpretation of the specific representant-choice (from the equivalence classes of $\mathbb{R}^d$) is not yet fully explained. On the other hand, this way the two sides (form and force diagram) of the duality-based construction have the same algebraic idea.

## 7. Summary

A dimension independent algebraic description of stress functions of graphic statics was given, containing previous descriptions in the respective dimensions. A graphic construction was also shown how the corresponding projective geometrical duality between two convex polyhedra and a pair of projections can be used to get a $d$ dimensional truss geometry and force diagram using Minkowski's theorem. It was observed how both diagrams can be treated as a projection of a drawing of a specific choice of homogenous point or hyperplane coordinates, in keeping with the $d$ dimensional algebraic description.

## References

[1] Baranyai T.: Analytical graphic statics. *International Journal of Space Structures*, to appear

[2] Cremona L. *Two treatises on the graphical calculus and reciprocal figures in graphical statics* (trans. TH Beare), Claredon Press, Oxford, 1890.

[3] Grünbaum B. and Shephard G.C. Duality of Polyhedra. In: *Shaping Space*, Senechal M. (eds) Springer, New York, NY. 2013, 211-216.

[4] Gunn C., *Geometry, Kinematics, and Rigid Body Mechanics in Cayley-Klein Geometries*. PhD thesis, Technical University Berlin, 2011. http://opus.kobv.de/tuberlin/volltexte/2011/3322.

[5] Hegedűs I., Stress function of single-layer reticulated shells and its relation to that of continuous membrane shells. *Acta Technica Academiae Scientiarum Hungaricae*, 1984; **97**; 103–110.

[6] Klain D.A., The Minkowski problem for polytopes. *Advances in Mathematics*, 2004; **185**; 270-288.

[7] Mars A., *Mathematics for Physics: An Illustrated Handbook*, World Scientific, 2017.

[8] Maxwell J.C. On Reciprocal Figures, Frames, and Diagrams of Forces, *Proceedings of the Royal Society of Edinburgh, 1872; 53-56.*

[9] McRobie A., The geometry of structural equilibrium. *Royal Society Open Science*, 2017; **4**

[10] McRobie A, Baker W, Mitchell T. et al., Mechanisms and states of self-stress of planar trusses using graphic statics, part ii: Applications and extensions. *International Journal of Space Structures,* 2016; **31**; 102–111.

[11] McRobie F.A. and Williams C.J.K.: A stress function for 3D frames, *International Journal of Solids and Structures,* 2017; **117**; 104–110.

[12] Mitchell T, Baker W, McRobie A. et al., Mechanisms and states of self-stress of planar trusses using graphic statics, part i: The fundamental theorem of linear algebra and the airy stress function. *International Journal of Space Structures*, 2016; **31;** 85–101.

[13] Phillips H.B., Stress functions. *Journal of Mathematics and Physics,* 1934; **13**; 421–425.

[14] Rankine WM. XVII. principle of the equilibrium of polyhedral frames. *The London, Edinburgh, and Dublin Philosophical Magazine and Journal of Science,* 1864; **27;** 92–92.

[15] Selig J.M., *Geometrical methods in Robotics*, Springer, 1996.

[16] Szőkefalvi N.GY., Gehér L., Nagy P., *Differenciálgeometria*, Műszaki Könyvkiadó, Budapest, 1979.

[17] Whiteley W. Motions and stresses of projected polyhedra. *Structural Topology*, 1982; **7;** 55-78.

[18] Williams C.J.K. and McRobie F.A., Graphic statics using discontinuous airy stress functions. *International Journal of Space Structures,* 2016; **31**; 121–134.